\def\final{0}
\documentclass[superscriptaddress, twocolumn, notitlepage, amsmath, amssymb, aps,
pra,
 10pt, letterpaper]{revtex4-1}

\usepackage{graphicx}% Include figure files
\usepackage{dcolumn}% Align table columns on decimal point
\usepackage{bm}% bold math

\usepackage{fancyhdr}
\usepackage{latexsym}
\usepackage{graphicx,color}
\usepackage{epstopdf}  %%after  \usepackage{graphicx,color}
\usepackage[pdfstartview=FitH]{hyperref}
\usepackage{xargs}
\usepackage{shadow,epsf,amsthm,amssymb,amsmath}

\usepackage{enumerate}

\usepackage{setspace}                           %for doublespacing
 
\usepackage{lipsum}% just to automatically generate text

\usepackage{tikz}	
\usetikzlibrary{backgrounds,fit,decorations.pathreplacing}  % TikZ libraries

\usepackage[colorinlistoftodos,prependcaption]{todonotes}

\usepackage{subfigure}

\newtheorem{definitionenv}{Definition}
\newtheorem{lemmaenv}[definitionenv]{Lemma}
\newtheorem{theoremenv}[definitionenv]{Theorem}
\newtheorem{corollaryenv}[definitionenv]{Corollary}
\newtheorem{propositionenv}[definitionenv]{Proposition}
\newtheorem{conjectureenv}[definitionenv]{Conjecture}
\newtheorem{remarkenv}[definitionenv]{Remark}
\newenvironment{remark}{\begin{remarkenv}\rm}{\end{remarkenv}}
\newcommand{\br}{\begin{remark}}
\newcommand{\er}{\end{remark}}

\newtheorem{exampleenv}{Example}
\newtheorem{app-lemmaenv}[section]{Lemma}

\newenvironment{definition}{\begin{definitionenv}\rm}{\end{definitionenv}}
\newenvironment{lemma}{\begin{lemmaenv}\rm}{\end{lemmaenv}}
\newenvironment{theorem}{\begin{theoremenv}\rm}{\end{theoremenv}}
\newenvironment{corollary}{\begin{corollaryenv}\rm}{\end{corollaryenv}}
\newenvironment{example}{\begin{exampleenv}\rm}{\end{exampleenv}}
\newenvironment{proposition}{\begin{propositionenv}\rm}{\end{propositionenv}}
\newenvironment{conjecture}{\begin{conjectureenv}\rm}{\end{conjectureenv}}
\newenvironment{app-lemma}{\begin{app-lemmaenv}\rm}{\end{app-lemmaenv}}

\newcommand{\bd}{\begin{definition}}
\newcommand{\ed}{\end{definition}}
\newcommand{\bl}{\begin{lemma}}
\newcommand{\el}{\end{lemma}}
\newcommand{\elp}{\hspace*{\fill} $\Box$
                 \end{lemma}}
\newcommand{\bt}{\begin{theorem}}
\newcommand{\et}{\end{theorem}}
\newcommand{\etp}{\hspace*{\fill} $\Box$
                 \end{theorem}}
\newcommand{\bc}{\begin{corollary}}
\newcommand{\ec}{\end{corollary}}
\newcommand{\ecp}{\hspace*{\fill} $\Box$
                 \end{corollary}}
\newcommand{\bcj}{\begin{conjecture}}
\newcommand{\ecj}{\end{conjecture}}

\newcommand{\be}{\begin{example}}
\newcommand{\ee}{\end{example}}
\newcommand{\eep}{\hspace*{\fill} $\Box$
                 \end{example}}
\newcommand{\bp}{\begin{proposition}}
\newcommand{\ep}{\end{proposition}}
\newcommand{\epp}{%\hspace*{\fill} $\Box$
                 \end{proposition}}

\newcommand{\eeq}{ \setcounter{equation} {\value{enumi}}}

\newcommand{\bfa}{{\mathbf a}}
\newcommand{\bfb}{{\mathbf b}}

\newcommand{\bfx}{{\mathbf x}}
\newcommand{\bfy}{{\mathbf y}}

\newcommand{\sdp}{{SD protocol}}

\renewcommand{\_}{\underline}

\def\beq{\begin{equation}}
\def\eeq{\end{equation}}

\def\bean{\begin{IEEEeqnarray*}{rCl}}
\def\eean{\end{IEEEeqnarray*}}

\ifnum\final=0
\newcommand{\mynote}[2]{{\color{#1} \marginpar{\tiny #2}}}
\newcommand{\mybignote}[2]{{\color{#1} $\langle \langle$ #2$\rangle \rangle$}}
\newcommandx{\rednote}[2][1=]{\todo[linecolor=red,backgroundcolor=red!25,bordercolor=red,#1]{#2}}
\newcommandx{\bluenote}[2][1=]{\todo[linecolor=blue,backgroundcolor=blue!25,bordercolor=blue,#1]{#2}}
\newcommandx{\yellownote}[2][1=]{\todo[linecolor=yellow,backgroundcolor=yellow!25,bordercolor=yellow,#1]{#2}}
\newcommandx{\greennote}[2][1=]{\todo[inline,linecolor=olive,backgroundcolor=green!25,bordercolor=olive,#1]{#2}}

\else
\newcommand{\mynote}[2]{}
\newcommand{\mybignote}[2]{}
\newcommand{\rednote}[2][1=]{}
\newcommand{\bluenote}[2][1=]{}
\newcommand{\greennote}[2][1=]{}
\newcommand{\yellownote}[2][1=]{}

\fi

\begin{document}

\preprint{APS/123-QED}

\title{Carrying an arbitrarily large amount of information using a 
	single quantum particle
}% Force line breaks with \\

\author{Li-Yi Hsu}
% \email{lyhsu@cycu.edu.tw}
 \affiliation{\footnotesize Department of Physics, Chung Yuan Christian University, Chungli 32023, Taiwan }

\author{Ching-Yi Lai}
\email{cylai@nctu.edu.tw}
\affiliation{\footnotesize Institute of Communications Engineering, National Chiao Tung University, Hsinchu 30010, Taiwan}

\author{You-Chia Chang}
 \affiliation{\footnotesize Department of Photonics and Institute of Electro-Optical Engineering, National Chiao Tung University, Hsinchu 30010, Taiwan}

 \author{Chien-Ming Wu}
 \affiliation{\footnotesize Institute of Photonics Technologies, National Tsing Hua University, Hsinchu 30013, Taiwan}

 \author{Ray-Kuang Lee}
 \affiliation{\footnotesize Institute of Photonics Technologies, National Tsing Hua University, Hsinchu 30013, Taiwan}

\date{\today}% It is always \today, today,

\begin{abstract}
Theoretically speaking, a photon can travel arbitrarily long before it enters into a detector, resulting a click. How much information can a photon carry? We study a bipartite asymmetric ``two-way signaling" protocol as an extension of that proposed by Del Santo and Daki\ifmmode \acute{c}\else \'{c}\fi{}. %(Phys. Rev. Lett. 120, 060503 (2018)). 
Suppose that Alice and Bob are distant from each other and each of them has an $n$-bit string. They are tasked to exchange the information of their local n-bit strings with each other, using only a single photon during the communication. It has been shown that the superposition of different spatial locations in a Mach-Zehnder (MZ) interferometer enables bipartite local encodings. We show that, after the travel of a photon through a cascade of $n$-level MZ interferometers in our protocol, the one of Alice or Bob whose detector clicks can access the other's full information of $n$-bit string,
while the other can gain one-bit of information. That is, the wave-particle duality makes two-way signaling possible, and a single photon can carry arbitrarily large (but finite) information.
\end{abstract}

\maketitle

\section{Introduction}
  Communication is a process of sending and receiving messages from one party to another~\cite{Shannon48}. More precisely, communication is a physical process with physical information carriers transmitted without violating any physical principle. For instance, electromagnetic waves used in wireless communication are governed by Maxwell's equations in classical physics. As a consequence of special relativity, faster-than-light communication is impossible. Also, the unavoidable energy consumption in the Maxwell's demon and Landauer's erasure indicates that information is physical~\cite{Lan91}, and the link between thermodynamics and information has potential to deliver new insights in physics and biology. 
 
 %Thank to quantum information science, 
 The role of information in physics theory has been extensively investigated. For example, it is proposed that quantum theory can be derived and reconstructed from purely informational principles~\cite{Har01,DB10,CDP11, MMA+13}. The effect of uncertainty relation in information processing can be stated in terms of information content principle~\cite{CHHH17} and No-Disturbance-Without-Uncertainty principle~\cite{SZY19}. Therein, a fundamental and interesting concern is the channel capacity in communication.  {According to the no-signaling principle, there is no information gain without classical or quantum communication;} the transmission of the message as the cause that increases the information. 
 %Eventually, using quantum particles as information carriers, there are many features in quantum communication distinguished from classical communication. 
 It is well known that, in the dense-coding protocol, two bits of information can be carried in one qubit with preshared entanglement~\cite{BW92}. For the receiver to obtain $n$ bits of information, at least a total of $n$ qubits have to be exchanged and at least $n/2$ qubits have to be sent from the sender~\cite{Hol73,CDNT99,DHW04,NS06,HW10,DH11,LC18c}. As a generalization of no-signaling principle {respected both in classical and quantum physics}, information causality states that one cannot gain more information than the number of bits sent via classical communication~\cite{PPK+09}. 
 %A random access code is employed as the task of information causality. 
 Note that the protocols mentioned above are proposed for one-way communication, and quantum entanglement as physical resource is initially distributed between the sender and receiver.

 Photons as flying qubits are usually exploited in quantum communication. 
 %Generating entangled photon pairs is much more difficult than preparing single-photon sources with near-term  photonic technology. 
 Given a photon as an information carrier, its particle-wave duality makes two-way communication possible. %(see the \textit{SD game} below).  
 Very recently  a variant of  the ``guess your neighbor's input" game~\cite{ABB+10} was studied by
 Santo and Daki\'{c}~\cite{SD18},  which we call  the SD game in this article.
 They proposed a protocol (\sdp) to win the SD game with certainty,
 while a classical strategy can win with probability at most $50\%$.
 We review the SD game as follows. Two distant agents Alice and Bob are given two input bits $x,y\in\{0,1\}$, respectively, which are drawn uniformly at random, and they are asked to output two bits $a,b\in\{0,1\}$, respectively. They win the game if   both of them  output a bit that is equal to the other's input (i.e., $a=y$ and $b=x$). {A restriction here is that only an information carrier, classical or quantum, can be manipulated. Obviously, 
 %one-bit classical communication must be one-way communication, and 
 they cannot win with certainty using simply a classical information carrier since it can transmit a single bit of information within a specified time limit.} Using a photon, on the other hand, enables two-way signalling so that they can win the SD game with certainty~\cite{MMD+18}. Notably, one of them can gain one bit of information if no detector clicks. 
 %(This protocol has been experimentally demonstrated~\cite{MMD+18}.)
 %Specifically, a 50/50 beam splitter can make a photon into two spatially-separated parts in a coherent way, which behaves wave-like. Two local agents, each of which holds a part of the photon, can encode information on the accessible light paths. On the other hand, the decoding process exploit the particle-like behavior of the photon. Let these two coherent parts meet at another beam splitter and then the interference determines which way the photon goes. 
 According to Renninger's negative result experiment~\cite{Ren53, Bae05} or the bomb-testing problem~\cite{Elitzur1993}, even if there is no interaction between the quantum object and the measuring device, one   still learns definite knowledge of the quantum state. %It is possible for a local agent to gain the other's information even though the detectors at hand do not click. That is, two-way signaling can be achieved using a single photon so that each of the local agents can obtain non-zero information gain. 
 
 %\rmark{DELETE Classical communication is forbidden during the game; they are allowed to exploit only a single photon as the information carrier. For example, Alice can imprint her input $x$ into the phase of the photon and then send it to Bob. In this way, they win the game with probability no larger than $1/2$ since Alice has to guess Bob's input $y$~\cite{SD18}.DELETE}
 
\begin{figure}[h!]
 \centering
   \includegraphics[width=0.5\linewidth] {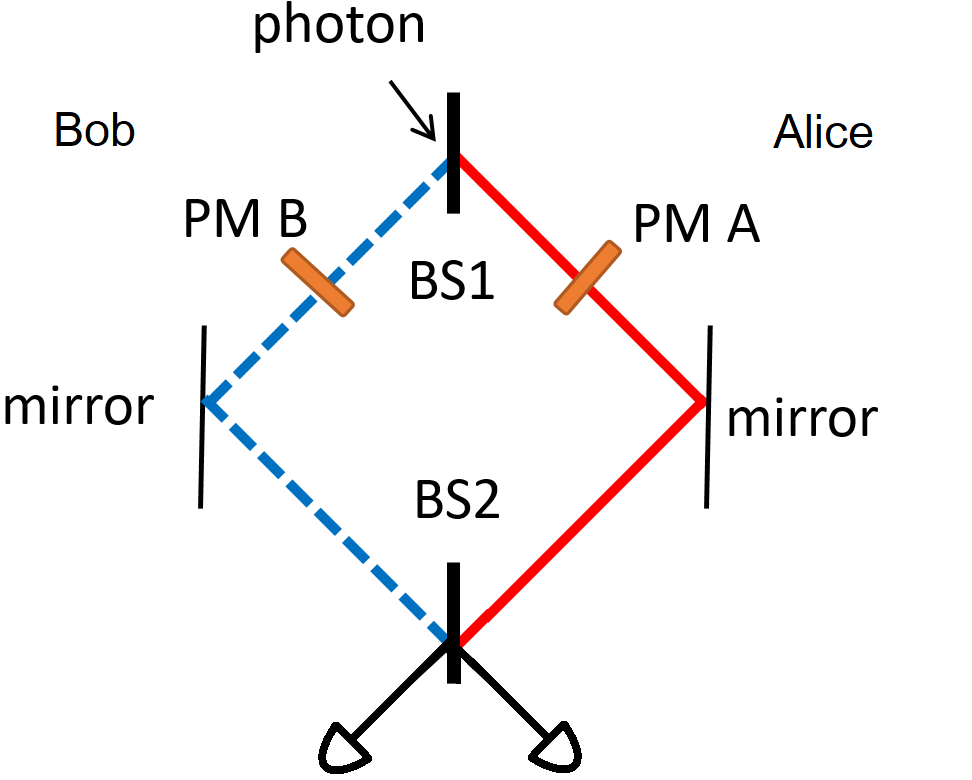} 
 \caption{ Optical implementation of the SD protocol.
  The photon is initially injected into the level-1 MZ interferometer. After traveling through the first $50/50$ beam splitter (BS1), the incident photon is half-reflected and half-transmitted in a coherent way. Alice and Bob can locally access the halves depicted in the red solid line and blue dash line, respectively.  For the local encoding at level~1, Alice inserts the $\pi$-phase modulator (PM A) if $x_{1}=0$ and does nothing if  $x_{1}=1$; similarly, Bob inserts the $\pi$-phase modulators, (PM B) if $y_{1}=0$ and does nothing if $y_{1}=1$. After the interference of these two coherent halves  at the second $50/50$ beam splitter (BS2), the photon enters one of the two detectors~\cite{SD18}.} \label{fig:SD_game}
\end{figure}

 %{As previously mentioned, a photon can be exploited in two-way communication.} The {\sdp} revealed a photon's power of ``two-way signaling"~\cite{SD18}, 
% and this protocol has been experimentally demonstrated~\cite{MMD+18}. 
 The concept of {\sdp} is explained as follows, in terms of the (level 1) optical implementation shown in~Fig.~\ref{fig:SD_game}.
A photon is emitted from a referee source and then injected into the first beam splitter (BS 1) of a Mach-Zehnder (MZ) interferometer. Consequently, this single photon is coherently superposed over  two different spatial locations. Hence the two local agents Alice and Bob can each (i) perform local operations on the incoming parts of the photon as information encoding, and (ii) access a detector to detect the photon at a certain time window later.  %\bmark{(see Fig.~\ref{fig:SD_game}(a))\cnote{Check the Fig. 1 (a) or Fig. 1}}. 
    According to (i), Alice and Bob encode their bits in the phase of the photon before it reaches the second beam splitter (BS2). With a delicate design, the parity of the two input bits completely determines the path of the photon leaving BS2. Consequently,  one knows with certainty which detector will detect  this photon while the other will detect nothing. For example, Alice's (Bob's) detector clicks if $x=y$ $(x\neq y)$ in the ideal case. %In this case, 
    Once Alice's detector does not receive any photon in a certain time window (no interaction between the quantum object and the measuring device), she knows that $x\neq y$ and outputs bit $a=x+1 \mod  2$. 
{As a result, using the spacial superposition of a single photon,  Alice and Bob can communicate a total of two-bit information  within a specified time window and hence win the game with certainty. 
 As opposed to this quantum communication, to win the game with certainty using classical communication, the time-window would have been too short to exchange two one-way classical communications~\cite{SD18}.}
 %{either Alice or Bob can send one information carrier to the other within a specified time window. As a result, two-way quantum communication is allowed within a time-window that would have been too short to exchange two one-way classical communications ~
 %To win the game with certainty via classical communication, either Alice or Bob can send one information carrier to the other within a specified time window.  As a result, two-way quantum communication is allowed within a time-window that would have been too short to exchange two one-way classical communications~\cite{SD18}.
   
   % Alice and Bob each must send a classical bit to the other.Also the distance between Alice and Bob is chosen so that either Alice or Bob can send one information carrier to the other within a specified time window, but it is impossible for the carrier to get back to the sender in time.%
   
    %    \rmark{Later it will be shown that the maximal   time window to ensure two-way communication are too short to exchange classical one way communication.} 
    
%    \rmark{(As the end of this paragraph, information causality is proposed in the scenario where quantum entanglement as no-signaling resource is initially distributed. To win the game with certainty, Alice and Bob each must send a classical bit to the other. Therefore, a lesson learned from this game is that a photon as two-way signaling resource can be equivalent to two-bit, two-way classical communication with no-signaling resource. %(這樣寫好不好?). 
%    )}
    
    In this paper, we characterize the power of a single photon as an information carrier. Our concerns are twofold: how much information a single photon can carry; and how much information an agent can obtain even if an interaction-free measurement occurs (no photon is detected by the detectors at hand). 
    {We will design a generalized Santo and Daki\'{c} (GSD) game and show that using one single photon, one can win the game with certainty and learn a total of $(n+1)$ bits of information in an $n$-level GSD game, while one learns only $n$ bits of information by classical communication.
    When $n$ is arbitrarily large, this suggests that a single photon can carry an arbitrarily large amount of information. 
We would like to mention that in a related work~\cite{HD19}, Horvat and Daki\'{c} showed that a single particle can be used to communicate simultaneously with $n$ parties and achieves the so-called genuine $n$-way signaling.} 
Note that a photon as an information carrier here can be replaced by a quantum particle whose coherence is under enough experimental control to exhibit coherence.

The remainder of this paper is organized as follows: In Sec.~\ref{sec:GSD}, we introduce the GSD game. The experiment setup of the $n$-level circuit is proposed. We characterize and then optimize the total information gains for Alice and Bob. Several specific cases are  studied. In Sec.~\ref{sec:discussion}, we investigate the physics concerning the information gains. Finally, in Sec.~\ref{sec:implementation} we estimate the performance of the GSD game in the physical realization. 

\section{Generalized SD game}\label{sec:GSD}
\subsection{Experimental setup}
    We consider a generalized Santo and Daki\'{c} (GSD) game as follows. Alice and Bob are assigned two independent   input strings $\bfx=x_1  \cdots x_{n},$ $\bfy=y_1  \cdots y_{n}\in\{0,1\}^n$, respectively, and they are asked to output bit strings $\bfa=a_1 \cdots a_{n}$ and $\bfb=b_1 \cdots b_{n}$, respectively.  {They win the game if (i) one of them can know the other's input string and (ii) the other can gain at least one bit of information. Only a single information carrier is allowed for the communication task within a specific time window.}
    %They are allowed to communicate by using a single information carrier back and forth for a total of $n$ times.%
    %The referee is restricted to emit only a single photon flying to a MZ interferometer.
    %\rmark{time limit}
    %, so that $x=a$ and $y=b$. 
    %Without classical communication, 
    %They win the game if $a_{i}=y_{i}$ and $b_{i}=x_{i}$ for all\,$i$.% 
    %Intuitively, if the information carrier is classical, they can exchange a total of $n$ bits of information.%
    %If $n$ photons are available, they can repeat the {\sdp}   $n$ times and exchange $2n$ bits of information  to win the game with certainty. %z
    %\rmark{DELETE Thus the restriction of a single-photon transmission in this GSD game is important.DELETE}  
    Equipped with a single photon in the GSD game, it will be shown that %there can be non-zero information gain for both Alice and Bob and % 
   {there is} a total of $n+1$ bits of information gain   {for Alice and Bob} as a result of two-way signaling  {in a time window $\tau$.} {However, if the information carrier is classical, they can exchange a total of at most $n$ bits of information in the same time window.}
    %In particular, there is non-zero probability to win the game if one of them can access only a detector.%
    %\cnote{can we win with certainty? Only when Alice's detector clicks.}%
      \begin{figure}[h]
 \centering
   \includegraphics[width=1.0\linewidth] {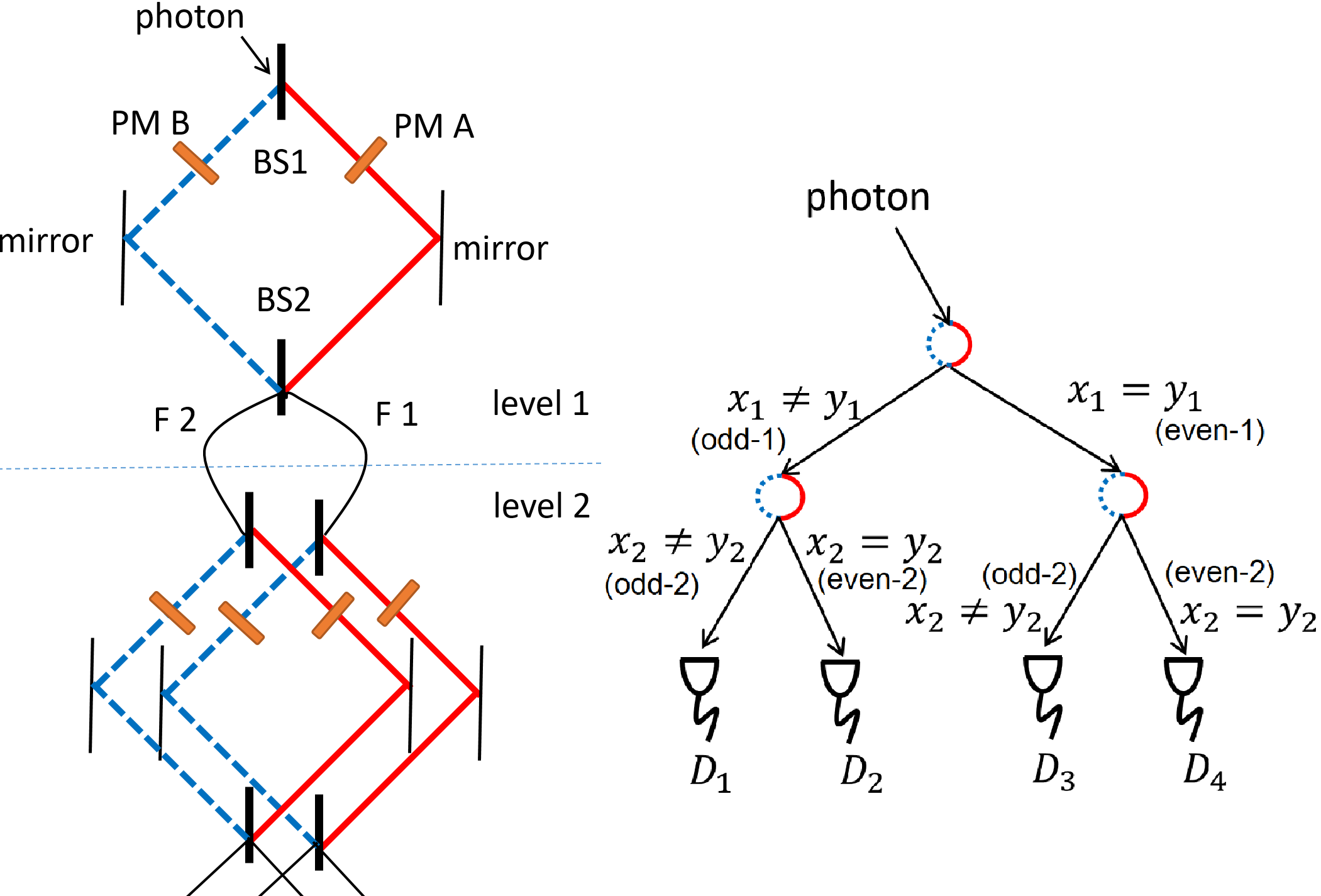} 
 \caption{Left: The optical details of a two-level evolution circuit.
 %The photon’s evolution of a 2-level circuit. 
 According to the local encoding at level~1, the leaving photon at BS2 is injected into one of the two fibers (F1 and F2), and enters one of these two MZ interferometers at level~2. Similarly, Alice (Bob) inserts PMs into these two MZ interferometers at level~2 if $x_{2}=1$ $(y_{2}=1)$ and does nothing if $x_{2}=0$ $(y_{2}=0)$. 
 %Here the bit values $x_{1}$, $y_{1}$, $x_{2}$, and $y_{2}$ are all set to $1$.
 Right: The topological unfolding of the 2-level circuit as a full 2-level binary tree. The nodes therein denote the MZ interferometers, where the photon is spatially superposed, and the directed edges between nodes indicates the possible travelling paths of the photon.
 } \label{fig:SD_game_l2}
\end{figure}

   First consider a two-level {circuit as the} extension of the SD protocol, as shown in Fig.~\ref{fig:SD_game_l2}.
The two detectors in Fig.~\ref{fig:SD_game} are replaced by MZ interferometers, followed by four detectors. 
One of the four detectors will click according to the parities of $(x_1,y_1)$ and $(x_2, y_2)$. 
%This can be naturally extended to an $n$-level scheme in the same way.%
A $(k+1)$-level circuit can be constructed by (i) replacing the detectors  in the $k$-level circuit by the MZ interferometers, and (ii) putting $2^{k+1}$ detectors at the output of the interferometers. {{Naturally}, an $n$-level circuit for GSD game can be {recursively} extended.}

\begin{figure}[!h]
 \centering
   \includegraphics[width=0.950\linewidth] {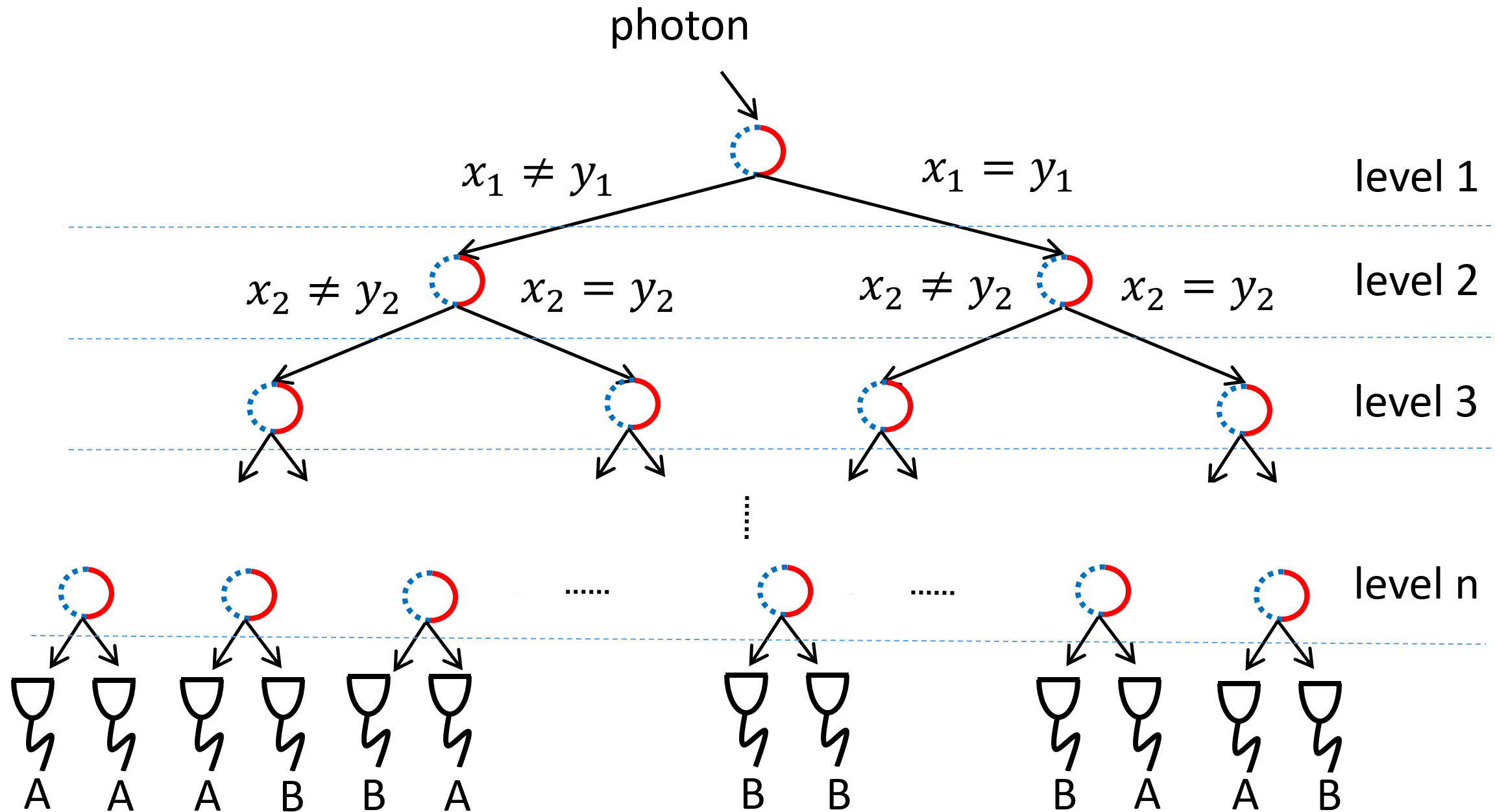}
\caption{ 
The unfolding layout of the $n$-level circuit as a perfect $n$-level binary tree and the detectors. There are $2^{i-1}$ MZ interferometers at level $i$. The photon is initially injected into the level-1 MZ interferometer. Then the parity of the bit pair $(x_{1}, y_{1})$ completely determines which one of the two level-2 MZ interferometers the photon will enter. Without loss of generality,  let the photon go to the right MZ interferometer at level 2 if $x_{1} = y_{1}$, and  the left MZ interferometer, otherwise. More optical details are explained in Fig.\,\ref{fig:SD_game}. Similarly, the parity of $(x_{2}, y_{2})$ determines the next target interferometers at level 3. This process is continued for a cascade of $n$ MZ interferometers. As a result, the light path of the photon completely depends on the $n$ bit pairs $(x_{1}, y_{1})$, \dots, $(x_{n}, y_{n})$. Finally, the photon flies into one of the $2^{n}$ detectors, each of which is held by Alice (A) or Bob (B).
 } \label{fig:1b}
\end{figure}

 Our protocol for the SGD game is explained as follows with the experimental setup 
 shown in Fig.~\ref{fig:1b}, which can be schematically depicted as a perfect $n$-level binary tree.
 A detector is placed at each leaf node, and a MZ interferometer is placed at each parent node. 
According to the input bits $x_{i}$ and $y_{i}$, Alice and Bob perform phase encoding by inserting a phase modulator (bit value $=0$) or not (bit value $=1$) into each of the $2^{i}$ MZ interferometers at level $i$. 
    Hence a single photon injected into the root will travel through one of the $2^n$ light paths. 
    Therein, after leaving a MZ interferometer at level $k$, the photon goes either the \textit{even-$k$} $(x_{k}=y_{k})$ path or \textit{odd-$k$} $(x_{k}\neq y_{k})$ one, and then enters into a MZ interferometer at level $(k+1)$. Note that there are $2^{k-1}$ even-$k$ and $2^{k-1}$ odd-$k$ paths. 
    %In Fig.~\ref{fig:1b} with k=1 as an example, the even-1 (odd-1) path is the right (left) one. 
    Consequently, the photon’s complete path is determined by the parity relations of the $n$ bit pairs $(x_{1}, y_{1}),$ \dots, $(x_{n}=y_{n})$ and finally flies into one of $2^{n}$ detectors, $D_{1},$ \dots, $D_{2^n}$, which are locally  accessible to either Alice or Bob. (Note that it is not necessary that Alice and Bob have an equal number of detectors.) The local agent whose detector clicks can learn these $n$ parities and hence knows the other's $n$ input bits exactly.

    %The light path taken after leaving the MZ interferometer at level $i$ is completely determined by the parity of $x_i$ and $y_i$, and then flies to one of the two following MZ interferometers at level $(i+1)$. After traveling through $n$ MZ interferometers in sequence, the photon finally goes to one of the $2^n$ detectors, which are locally accessible to either Alice or Bob. (Note that it is not necessary that Alice and Bob have an equal number of detectors.) 
    %The detector where the photon flies is completely determined by the bitwise parities of $\bfx$ and $\bfy$. 
    
\begin{figure}[!h]
 \centering
   \includegraphics[width=1.03\linewidth] {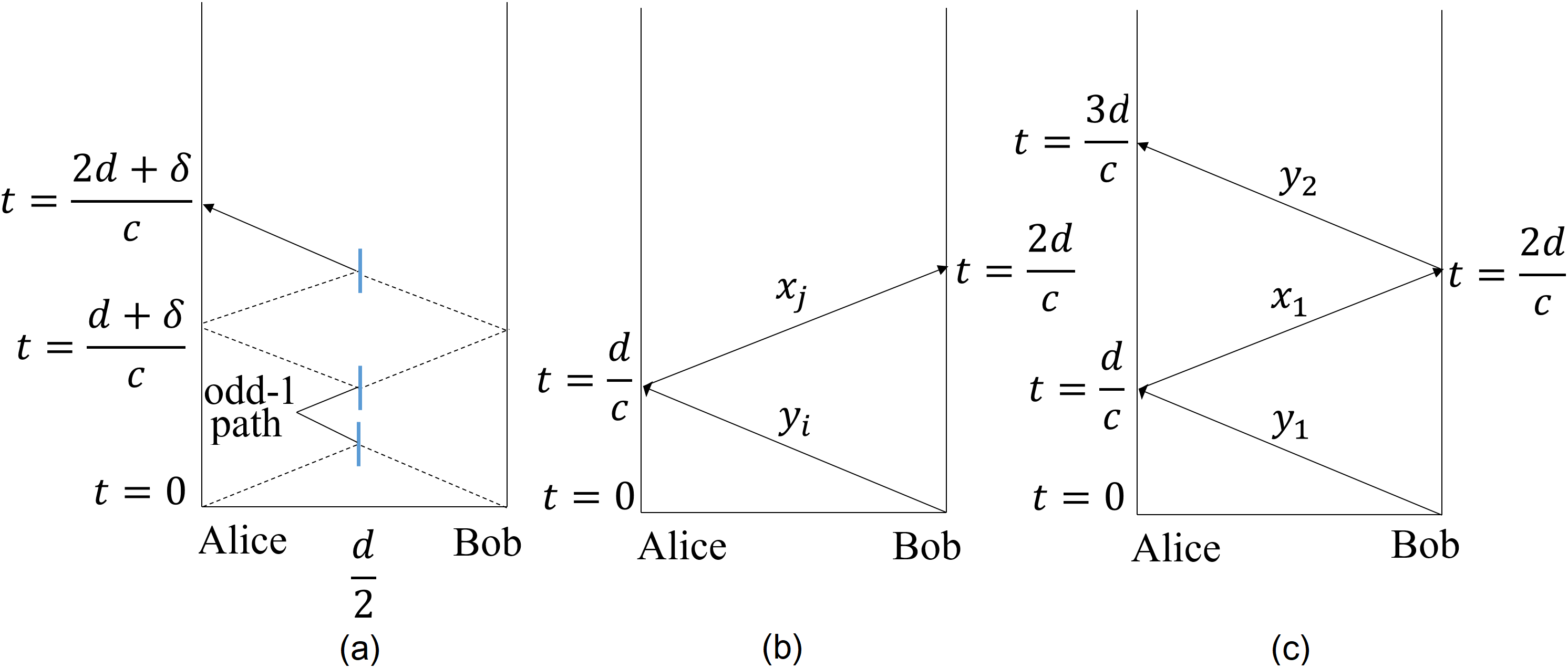}
\caption{ 
 {
 The time window in the $n=2$ case. (a) Assume that Alice holds the left two detectors ($D_1$ and $D_2$ in Fig.~\ref{fig:SD_game_l2}). Alice and Bob each performs the local encoding operation for $x_{1}$ at $t=0$  and for $y_{1}$ at $t=(d+\delta)/c$, respectively. Finally, the photon flies into one of the detectors accessible to Alice at time $(2d+\delta)/c\leq \tau\leq  (2d+\delta)/c+\epsilon$. (b) In the same window $\tau$, Alice and Bob can only exchange some $x_{j}$ and $y_{i}$ by classical communication. (c) To finish the same task as in (a) using  a classical information carrier, it requires a time window $3d/c \leq \tau' \leq 3d/c +\epsilon$.  }
 } \label{fig:time}
\end{figure}

 {
Next we discuss the physical settings so that Alice and Bob can exchange a single information carrier for a total of $n$ times. Let Alice and Bob be located at a distance $d$ from each other, and, for simplicity, assume that an information carrier, classical or quantum, travels at the speed $c$. 
Suppose that  an information carrier  carries one bit of information in classical one-way communication. So it takes time roughly $nd/c$ for transmitting $n$ bits of information by a single carrier.
On the other hand, let the length of an odd-$k$ or even-$k$ path be $\delta$ {for $k<n$}.
 {In other words, in Fig.~\ref{fig:SD_game_l2}, 
 %the length of the light path%
 {the photon travels a distance $d$ between BS1 and BS2 is, and the length of F1 or F2 is $\delta$.}
%We may assume that%
{In the experiment setup, let} $\delta\ll d$ by choosing sufficiently large $d$; however, 
%this assumption% 
{such setup} is not reflected from the scale of our plots.}
Thus it takes time $((nd+(n-1)\delta)/c)$ to implement our protocol in Fig.~\ref{fig:1b}. 
As a result, we allow a specific time window $\tau$ such that $((nd+(n-1)\delta)/c)\leq \tau \leq ((nd+(n-1)\delta)/c)+\epsilon$, where $\epsilon\geq 0$ is a small constant
such that $((nd+(n-1)\delta)/c)+\epsilon< (n+1)d/c$.
This choice of time window $\tau$ allows Alice and Bob to exchange a total of $n+1$ bits of information (shown in the next subsection)  using our protocol, but this time window is not long enough so that  a classical scheme can  exchange only $n$ bits of information.
}
An example of $n=2$ is illustrated in Fig.~\ref{fig:time}.

  \begin{table*}[t!]
           \centering
           \begin{tabular}{|c|c|c|c|c|c|c|}
           \hline
           &$D_1$& $D_2$& $D_3$& $D_4$& Bob's knowledge on the bit-pair relations& Bob's information gain \\
           \hline
            Case (1) &A&A&B&B& $x_1\neq y_1$  &1\\
             \hline
            Case (2) &B&B&A&A& $x_1= y_1$&1  \\
             \hline
            Case (3) &A&B&A&B& $x_2\neq y_2$&1  \\
             \hline
            Case (4) &B&A&B&A& $x_2= y_2$ &1 \\
             \hline
            Case (5) &B&A&A&B& either $x_1\neq y_1$ and  $x_2\neq y_2$, or $x_1=y_1$ and $x_2=y_2$ &1 \\
            \hline
            Case (6) &A&B&B&A& either $x_1\neq y_1$ and  $x_2= y_2$, or $x_1=y_1$ and $x_2\neq y_2$&1  \\
            \hline
             Case (7) &A&B&B&B& $x_1\neq y_1$ and $x_2\neq y_2$  & $2$ \\
             \hline
            Case (8) &B&A&A&A& $(x_1,x_2)\neq (y_1,y_2)$  & $2-\log_2 3$ \\
            \hline
           \end{tabular}
           \caption{Some detector assignments and Bob's corresponding information gains, given that one of Alice's detectors clicks. In the cases (1) and (2) ((3) and (4)), Bob knows $x_1 (x_2)$ with certainty. In the cases (5) and (6), Bob knows $x_1$ and $x_2$ simultaneously with probability 0.5, which indicates that Bob can gain one-bit information on average. In case (7) Bob knows that
           $x_1\neq y_1$ and $x_2\neq y_2$ and his information gain is $2$ (bits).
           In case (8), Bob knows that $(x_1,x_2)\neq (y_1,y_2)$ and hence his information gain is $2-\log_2 3$. }
           \label{tb:1}
       \end{table*}

%(ii) We add Fig~\ref{fig:time} for illustrating the maximal time window in the $n=2$ case. 
%\begin{figure}[h!]
% \centering
%   \includegraphics[width=0.9\linewidth] {Fig3.png} 
% \caption{ Left: 
% Right: Suppose that Alice and Bob each performs their local encoding initially at time $t=0$. 
% } \label{fig:time}
%\end{figure}

\begin{figure}[h!]
 \centering
   \includegraphics[width=0.99\linewidth] {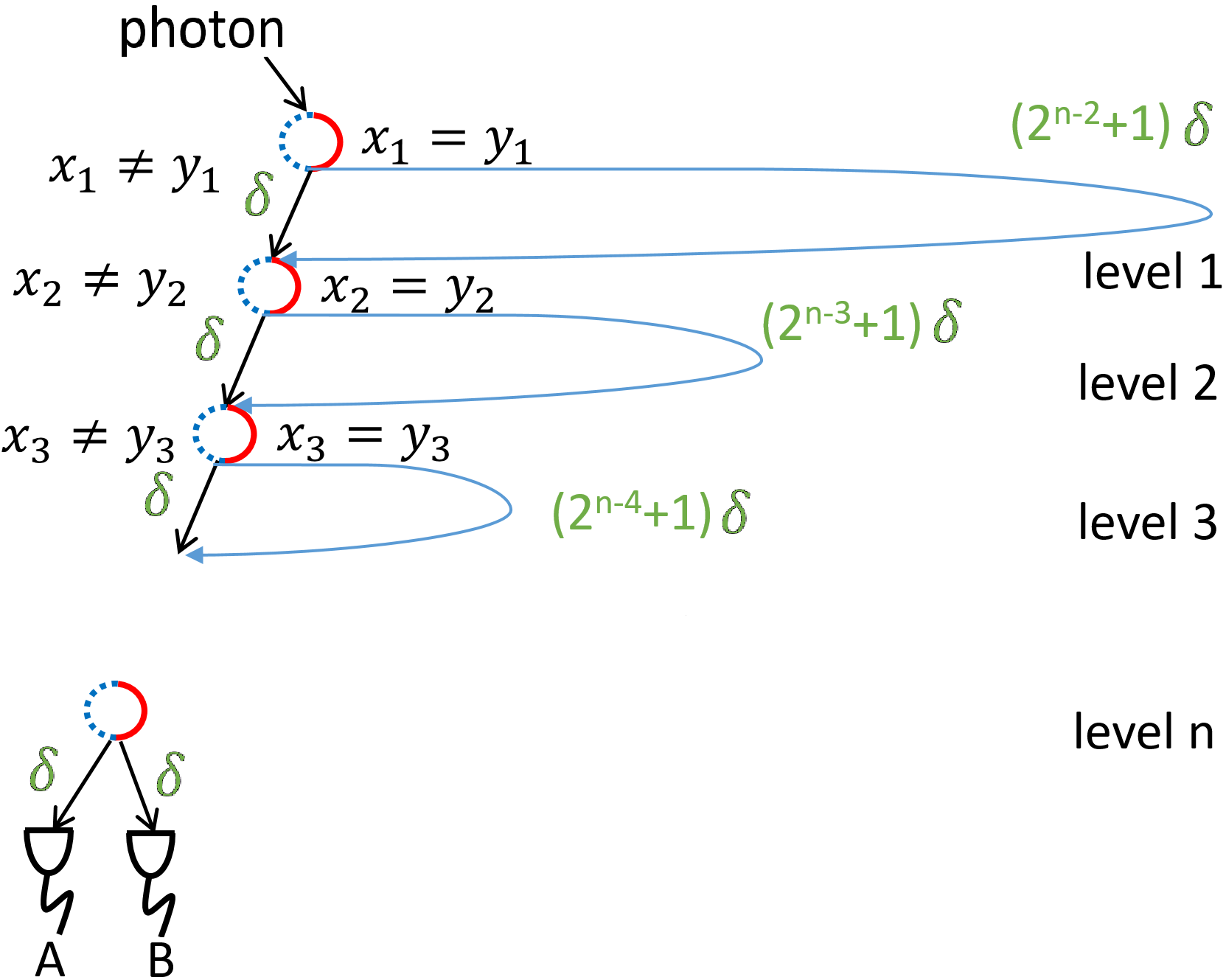} 
 \caption{An effective circuit with two detectors. All the left paths have a time delay $\delta$, while the right paths have additional delays denoted by longer optical fibers. 
 } \label{fig:2detectors}
\end{figure}

The implementation of Fig.~\ref{fig:1b} can be refined in the case that the detectors at the left and right leaves belong to Alice and Bob, respectively,
assuming that $2^n\delta \ll d$. {That is, $n$ cannot be arbitrarily large or $n=O(\log(d/\delta ))$.}
Specifically, we can use only two detectors (one for Alice and the other for Bob) and add a time domain coordinate to save the massive number of $2^n$ detectors required. 
This is done as shown in Fig.~\ref{fig:2detectors},
where the left path at level $i$ has a time delay $\delta$ and the right light path at level $i$ has an additional delay of $2^{n-i-1}\delta$ and both the left and right paths the level $n$ have delay $\delta$. 
Consequently, 
the $2^n$ light paths  from left to right in Fig.~\ref{fig:1b} will have delays $n\delta, n\delta, (n+1)\delta, (n+1)\delta, \dots, (2^{n-1}-1)\delta, (2^{n-1}-1)\delta$ in Fig.~\ref{fig:2detectors}, respectively. 
An important observation is that at the same level each of Alice and Bob has the same input bit
and applies the same PMs to the corresponding light paths. Also a beam splitter has two input ports, which allows us to connect both branches to the same beam splitter. 
Therefore, from the time of clicking, one can deduct the corresponding light path in the circuit of Fig.~\ref{fig:1b} and learn the $n$-bit string.

As a comparison, the previous scheme by Santo and Daki\'{c}~\cite{SD18} uses one single-photon source and two detectors to exchange two bits of information.
Our scheme is able to transmit more information at the cost of additional fibers, MZ interferometers, and beam splitters.  

\subsection{Information gains}

Let us quantify how many detectors Bob should have to optimize his information gain $I(X;B|Y)$, where $I(X;B|Y)= H(B|Y)-H(B|X,Y)$ is the \textit{mutual information} between Alice's input variable $X$ and Bob's output variable $B$ conditioned on Bob's input variable $Y$; $H(B|Y)$ is the \textit{conditional Shannon entropy}, and $H(X)= -\sum_{\bfx} p_\bfx \log p_\bfx$. Let $m$ be the number of detectors that belong to Bob.  Since $X$ and $Y$ are independent, it is clear that 
\begin{align*}
I(X;B|Y)\leq & H(B) =H\left(\left\{ \underbrace{{1}/{2^n},\dots, {1}/{2^n}}_m,1- {m}/{2^n}\right\}\right)\\
=&n-\left(1-\frac{m}{2^n}\right)\log\left(2^n-m\right).
\end{align*}
The total information gain of Alice and Bob is  
\begin{align*}
&I(Y;A|X)+I(X;B|Y) =H(A)+ H(B)\\ 
&=2n-\frac{m}{2^n}\log m- \left(1-\frac{m}{2^n}\right)\log\left(2^n-m\right)
\leq n+1,
\end{align*}
where the equality holds when $m=2^{n-1}$. The main result can be stated as follows:

%As a consequence, Alice and Bob should have an equal number of detectors to optimize their information gain. 
      { 
      \textit{The optimal total information gain is $n+1$.}
            }
       
To reach optimal total information gain, Alice and Bob each should access half of the $2^n$ detectors. It does not matter which detectors Alice or Bob should hold since the one with a clicking detector  can learn $n$ bits of information, while the other learns one bit of information.
     As an illustration,  we analyze Bob's information gain in the case of $n=2$
       as shown in Fig.~\ref{fig:SD_game_l2}. Various detector assignments as listed in Table~\ref{tb:1}, 
       assuming that one of Alice's (Bob's) detectors always clicks (never click).

       %In this case, Alice knows $y$ and hence gains $n$-bit of information. 
  %On the other hand, even though there is no interaction between Bob's measuring devices and the photon, he still can obtain   one-bit of information.       %\rmark{\textit{Main result.}} The optimal total information gain for Alice and Bob is $(n+1)$ bits <cite>SM</cite>.     %In other words, a single photon can carry infinite information as $n$ goes to infinity. 

     %In addition, what  local information can be learned is closely related to the local accessibility of these $2^n$ detectors between Alice and Bob. As a concrete example, suppose that Alice and Bob can access left and right halves of the detectors shown in Fig.~\ref{fig:1b}, respectively. Since none of Bob's detectors clicks, he knows the relation $x_1= y_1$ must be false and hence learns the bit $b_1=y_1+1 \mod 2$. 
     
{Note that, to win the GSD game with certainty, the one that cannot learn the other's $n$ input bits must know one bit of information.} 
     With a delicate initial assignment of these $2^n$ detectors between Alice and Bob, they can exchange the input bit pair $(x_{k}, y_{k})$ with certainty for some specific~$k$.
     Specifically, denote two detector sets by $\Delta_{k}^{E}$ and $\Delta_{k}^{O}$.  For all $i=1,\dots ,2^n$, the detector $D_{i}\in\Delta_{k}^{E}$ $(D_{i}\in \Delta_{k}^{O})$ if $D_{i}$ receives a photon travelling through an even-k (odd-k) path. Let all $2^{n-1}$ detector elements in $\Delta_{k}^{E}$ $(\Delta_{k}^{O})$ are completely accessible to Alice (Bob). In this case, once none of the detectors belonging to Bob clicks. Bob can learn that that $x_{k}=y_{k}$ and hence {he outputs the bit $b_{k}=x_{k}$ with certainty.}
     For example, as shown in Table~\ref{tb:1}, Bob can always learn $x_1$ using the detector assignment in Case (1) or (2), or learn $x_2$ using the detector assignment in Case (3) or (4).
%         \rmark{Finally, denote the detector set $\Delta_{(k)}^{E}$ ($\Delta_{(k)}^{O}$) and $D_{i}$ belongs to $\Delta_{(k)}^{E}({\Delta}_{(k)}^{O})$ if $D_{i}$ receives a photon traveling through an even-$k$ (odd-$k$) path. There are $2^{n-1}$ detector elements in the sets  ${\Delta}_{(k)}^{E}$ $\Delta_{(k)}^{O}$, respectively. }

   It is noteworthy to mention the following detector assignment. Let Alice occupy only one detector and Bob occupy the other $2^n-1$ ones. With probability $2^{-n}$, Alice's detector receives a photon. In this case, the no-click on Bob's side makes him exclude the possibility of $2^n-1$ parity relation sets and hence learn the input string $\bfx$.
   In other words, if Alice and Bob are tasked to output $\bfa=\bfy$ and $\bfb=\bfx$, respectively, in the GSD game, they can win the game with the probability $2^{-n}$. 
   {On the other hand, if one of Bob's detectors clicks, Alice still learns $n-\log_2(2^n-1)$ bits of information.} 
   %Therefore, Alice and Bob can win the GSD game with probability at least $2^{-n}$.%
   
   %As for the $n=1$ case, Alice and Bob can always exchange one-bit information~\cite{SD18}.
    %\cnote{why certainty? only if Alice's detector clicks, which occurs with probability $1/2^n$.} 

    \section{Discussion}\label{sec:discussion}A lesson learned from the dense coding is that sending one qubit is equivalent to sending two classical bits; another lesson from information causality is that, if there is no quantum communication, the information gain is equal to the amount of classical communication. Notably, the dense coding and random access code each  (i) are one-way communication, and (ii) exploit quantum entanglement as physical resource. To the best of our knowledge, the protocols of SD and GSD games are the first ones for two-way signaling quantum communication. Therein, the spatial coherent superposition and wave-particle duality can be regarded as physical resources. From the two-way signaling aspect, these two quantum properties of a photon are more beneficial than quantum entanglement. In the proposed two-way signaling protocol, sending a photon with an $n$-level circuit is equivalent to sending $n+1$ bits, where $n$ can be arbitrarily large. Which agent can obtain the other’s information depends on the local bit strings $\bfx$ and $\bfy$, and the pre-assignment of these $2^{n}$ detectors to Alice or Bob. {In any way, there is always a detector that clicks, which indicates either $I(Y;A|X)= n$ or $I(X;B|Y)= n$ must hold, and hence we can conclude that $n\leq I(Y;A|X) +I(X;B|Y)\leq n+1$.}
    %\rmark{DELETE In particular, Alice and Bob can pre-agree to exchange the bits $x_{k}$ and $y_{k}$ with the deliberated design of the circuit.DELETE}
    
    From the causal perspective, the optimal information gain in the GSD game can be explained in a two-fold way. Firstly, an information carrier is consumed therein. Notably, regarding the classical communication, information causality states that the information gain cannot exceed the amount of classical communication. Thus sending-and-receiving a photon can result in one-bit information gain. Secondly, the two distant local operations at the same level fully determines into which way the photon enters in the next level, and this contributes the one bit information. In other words, only when the coherent superposed parts meet at BS2 {of a MZ interferometer in every level} as shown in Fig.~\ref{fig:1b}, the which-way uncertainty between these two beam splitters in the MZ interferometer vanishes, and consequently produces one-bit information. That is, a level contributes one-bit information gain. At the end, at most $(n+1)$ bits of information can be generated during a photon entering an $n$-level circuit. 
    
    For example, in the Elitzur-Vaidman bomb tester, a single photon is emitted, but one of its coherent parts is blocked and there is no interference at the second beam splitter of a MZ interferometer~\cite{Elitzur1993}. In this case, only a bit of information (whether the bomb explodes) is accessible. On the other hand, in the simple one-way SD game, assume that the bit $y_{1}=1$ is public, and the bit $x_{1}$ is unknown to Bob~\cite{SD18}. To inform Bob, Alice performs local operations on the accessible coherent superposed part. It is the interference at the second beam splitter brings Bob the bit value of $x_{1}$. 
    
\section{Implementation}\label{sec:implementation} 

Since the complexity of the $n$-level circuit grows exponentially in $n$ (or linearly in $n$  if the scheme of Fig.~\ref{fig:2detectors} is used), it is impossible to realize the optical circuit for arbitrarily large $n$ with imperfect devices. Noisy components, such as the photon source, beam splitters and detectors, will cause the photon to decay and hence limit the possible circuit level. 

\begin{figure}[h]
 \centering
   \includegraphics[width=0.8\linewidth] {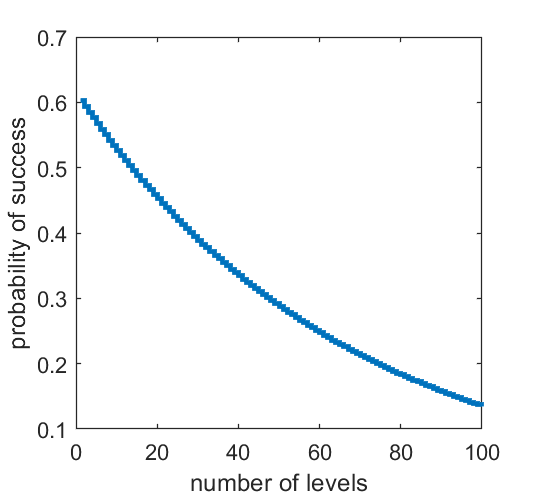} 
 \caption{The rate of success of our GSD protocol versus the number of circuit levels.
 } \label{fig:loss-rate}
\end{figure}

Here we estimate the performance of the protocol when it is implemented under realistic experimental conditions. We consider the following error sources. A realistic pulsed single-photon source has a photon number probability $P(n)$ to generate $n$ photons per pulse. A quantum dot single-photon source can achieve $P(1)=0.72$ \cite{SSW17}. The beam splitters in experiments may not have perfectly even split ratio between transmission and reflection, but this uneven split ratio can be compensated with experimental techniques, such as using wave plates together with polarized beam splitters. Therefore, we assume the split ratio is perfectly even. We also assume the phase errors given by phase shifter/modulators are negligible compared to other error sources. This is justifiable when using piezoelectric phase shifters, which can achieve a phase accuracy better than $2\pi /500$.  We consider the optical loss to be a dominating error source, which can result from the non-$100\%$ reflectivity of mirrors and the non-perfect anti-reflection coatings of all transmissive optical components.
We estimate the optical loss $\epsilon$ per stage to be $1.5\%$.  
%We estimate the optical loss $\epsilon$ to be $1.5$ % per stage [3]. %%
For example, using two AR coated surface for a wave plate and one AR coated surface for a beam splitter, each of $0.5\%$ loss. 
We assume the detection efficiency $\eta_D$ of the detectors to be $85\%$, which is achievable using superconducting nanowire single photon detectors (SNSPD). The contribution from the dark counts of the detectors can be negligible by using low-dark-count detectors such as SNSPD or by applying gating techniques. Using these numbers, we obtain the success rate of our protocol for $n$ stages to be $P(1)\left(1-\epsilon\right)^n \eta_D$, as shown in Fig.~\ref{fig:loss-rate}.

\emph{Acknowledgments.}|\,%We thank   for helpful discussions. 
CYL was supported from the Young Scholar Fellowship
Program by Ministry of Science and Technology (MOST) in
Taiwan, under Grant\,MOST108-2636-E-009-004.
YCC was supported by MOST108-2218-E-009-035-MY3.


\begin{thebibliography}{25}%
	\makeatletter
	\providecommand \@ifxundefined [1]{%
		\@ifx{#1\undefined}
	}%
	\providecommand \@ifnum [1]{%
		\ifnum #1\expandafter \@firstoftwo
		\else \expandafter \@secondoftwo
		\fi
	}%
	\providecommand \@ifx [1]{%
		\ifx #1\expandafter \@firstoftwo
		\else \expandafter \@secondoftwo
		\fi
	}%
	\providecommand \natexlab [1]{#1}%
	\providecommand \enquote  [1]{``#1''}%
	\providecommand \bibnamefont  [1]{#1}%
	\providecommand \bibfnamefont [1]{#1}%
	\providecommand \citenamefont [1]{#1}%
	\providecommand \href@noop [0]{\@secondoftwo}%
	\providecommand \href [0]{\begingroup \@sanitize@url \@href}%
	\providecommand \@href[1]{\@@startlink{#1}\@@href}%
	\providecommand \@@href[1]{\endgroup#1\@@endlink}%
	\providecommand \@sanitize@url [0]{\catcode `\\12\catcode `\$12\catcode
		`\&12\catcode `\#12\catcode `\^12\catcode `\_12\catcode `\%12\relax}%
	\providecommand \@@startlink[1]{}%
	\providecommand \@@endlink[0]{}%
	\providecommand \url  [0]{\begingroup\@sanitize@url \@url }%
	\providecommand \@url [1]{\endgroup\@href {#1}{\urlprefix }}%
	\providecommand \urlprefix  [0]{URL }%
	\providecommand \Eprint [0]{\href }%
	\providecommand \doibase [0]{http://dx.doi.org/}%
	\providecommand \selectlanguage [0]{\@gobble}%
	\providecommand \bibinfo  [0]{\@secondoftwo}%
	\providecommand \bibfield  [0]{\@secondoftwo}%
	\providecommand \translation [1]{[#1]}%
	\providecommand \BibitemOpen [0]{}%
	\providecommand \bibitemStop [0]{}%
	\providecommand \bibitemNoStop [0]{.\EOS\space}%
	\providecommand \EOS [0]{\spacefactor3000\relax}%
	\providecommand \BibitemShut  [1]{\csname bibitem#1\endcsname}%
	\let\auto@bib@innerbib\@empty
	%</preamble>
	\bibitem [{\citenamefont {Shannon}(1948)}]{Shannon48}%
	\BibitemOpen
	\bibfield  {author} {\bibinfo {author} {\bibfnamefont {C.~E.}\ \bibnamefont
			{Shannon}},\ }\href@noop {} {\bibfield  {journal} {\bibinfo  {journal} {Bell
				System Technical Journal}\ }\textbf {\bibinfo {volume} {27}},\ \bibinfo
		{pages} {379} (\bibinfo {year} {1948})}\BibitemShut {NoStop}%
	\bibitem [{\citenamefont {{Landauer}}(1991)}]{Lan91}%
	\BibitemOpen
	\bibfield  {author} {\bibinfo {author} {\bibfnamefont {R.}~\bibnamefont
			{{Landauer}}},\ }\href@noop {} {\bibfield  {journal} {\bibinfo  {journal}
			{Phys. Today}\ }\textbf {\bibinfo {volume} {44}},\ \bibinfo {pages} {23}
		(\bibinfo {year} {1991})}\BibitemShut {NoStop}%
	\bibitem [{\citenamefont {Hardy}(2001)}]{Har01}%
	\BibitemOpen
	\bibfield  {author} {\bibinfo {author} {\bibfnamefont {L.}~\bibnamefont
			{Hardy}},\ }\href@noop {} {\  (\bibinfo {year} {2001})},\ \Eprint
	{http://arxiv.org/abs/arXiv:quant-ph/0101012} {arXiv:quant-ph/0101012}
	\BibitemShut {NoStop}%
	\bibitem [{\citenamefont {Daki\'{c}}\ and\ \citenamefont {\v{C}.
			Brukner}(2011)}]{DB10}%
	\BibitemOpen
	\bibfield  {author} {\bibinfo {author} {\bibfnamefont {B.}~\bibnamefont
			{Daki\'{c}}}\ and\ \bibinfo {author} {\bibnamefont {\v{C}. Brukner}},\
	}\href@noop {} {\bibfield  {journal} {\bibinfo  {journal} {Deep Beauty:
				Understanding the Quantum World through Mathematical Innovation}\ ,\ \bibinfo
			{pages} {365}} (\bibinfo {year} {2011})},\ \Eprint
	{http://arxiv.org/abs/arXiv:0911.0695} {arXiv:0911.0695} \BibitemShut
	{NoStop}%
	\bibitem [{\citenamefont {Chiribella}\ \emph {et~al.}(2011)\citenamefont
		{Chiribella}, \citenamefont {D'Ariano},\ and\ \citenamefont
		{Perinotti}}]{CDP11}%
	\BibitemOpen
	\bibfield  {author} {\bibinfo {author} {\bibfnamefont {G.}~\bibnamefont
			{Chiribella}}, \bibinfo {author} {\bibfnamefont {G.~M.}\ \bibnamefont
			{D'Ariano}}, \ and\ \bibinfo {author} {\bibfnamefont {P.}~\bibnamefont
			{Perinotti}},\ }\href {\doibase 10.1103/PhysRevA.84.012311} {\bibfield
		{journal} {\bibinfo  {journal} {Phys. Rev. A}\ }\textbf {\bibinfo {volume}
			{84}},\ \bibinfo {pages} {012311} (\bibinfo {year} {2011})}\BibitemShut
	{NoStop}%
	\bibitem [{\citenamefont {Masanes}\ \emph {et~al.}(2013)\citenamefont
		{Masanes}, \citenamefont {M{\"u}ller}, \citenamefont {Augusiak},\ and\
		\citenamefont {P{\'e}rez-Garc{\'\i}a}}]{MMA+13}%
	\BibitemOpen
	\bibfield  {author} {\bibinfo {author} {\bibfnamefont {L.}~\bibnamefont
			{Masanes}}, \bibinfo {author} {\bibfnamefont {M.~P.}\ \bibnamefont
			{M{\"u}ller}}, \bibinfo {author} {\bibfnamefont {R.}~\bibnamefont
			{Augusiak}}, \ and\ \bibinfo {author} {\bibfnamefont {D.}~\bibnamefont
			{P{\'e}rez-Garc{\'\i}a}},\ }\href {\doibase 10.1073/pnas.1304884110}
	{\bibfield  {journal} {\bibinfo  {journal} {Proceedings of the National
				Academy of Sciences}\ }\textbf {\bibinfo {volume} {110}},\ \bibinfo {pages}
		{16373} (\bibinfo {year} {2013})},\ \Eprint
	{http://arxiv.org/abs/https://www.pnas.org/content/110/41/16373.full.pdf}
	{https://www.pnas.org/content/110/41/16373.full.pdf} \BibitemShut {NoStop}%
	\bibitem [{\citenamefont {Czekaj}\ \emph {et~al.}(2017)\citenamefont {Czekaj},
		\citenamefont {Horodecki}, \citenamefont {Horodecki},\ and\ \citenamefont
		{Horodecki}}]{CHHH17}%
	\BibitemOpen
	\bibfield  {author} {\bibinfo {author} {\bibfnamefont {L.}~\bibnamefont
			{Czekaj}}, \bibinfo {author} {\bibfnamefont {M.}~\bibnamefont {Horodecki}},
		\bibinfo {author} {\bibfnamefont {P.}~\bibnamefont {Horodecki}}, \ and\
		\bibinfo {author} {\bibfnamefont {R.}~\bibnamefont {Horodecki}},\ }\href
	{\doibase 10.1103/PhysRevA.95.022119} {\bibfield  {journal} {\bibinfo
			{journal} {Phys. Rev. A}\ }\textbf {\bibinfo {volume} {95}},\ \bibinfo
		{pages} {022119} (\bibinfo {year} {2017})}\BibitemShut {NoStop}%
	\bibitem [{\citenamefont {Sun}\ \emph {et~al.}(2019)\citenamefont {Sun},
		\citenamefont {Zhou},\ and\ \citenamefont {Yu}}]{SZY19}%
	\BibitemOpen
	\bibfield  {author} {\bibinfo {author} {\bibfnamefont {L.-L.}\ \bibnamefont
			{Sun}}, \bibinfo {author} {\bibfnamefont {X.}~\bibnamefont {Zhou}}, \ and\
		\bibinfo {author} {\bibfnamefont {S.}~\bibnamefont {Yu}},\ }\href@noop {} {\
		(\bibinfo {year} {2019})},\ \bibinfo {note} {arXiv:1906.11807}\BibitemShut
	{NoStop}%
	\bibitem [{\citenamefont {Bennett}\ and\ \citenamefont {Wiesner}(1992)}]{BW92}%
	\BibitemOpen
	\bibfield  {author} {\bibinfo {author} {\bibfnamefont {C.~H.}\ \bibnamefont
			{Bennett}}\ and\ \bibinfo {author} {\bibfnamefont {S.~J.}\ \bibnamefont
			{Wiesner}},\ }\href@noop {} {\bibfield  {journal} {\bibinfo  {journal} {Phys.
				Rev. Lett.}\ }\textbf {\bibinfo {volume} {69}},\ \bibinfo {pages} {2881}
		(\bibinfo {year} {1992})}\BibitemShut {NoStop}%
	\bibitem [{\citenamefont {Holevo}(1973)}]{Hol73}%
	\BibitemOpen
	\bibfield  {author} {\bibinfo {author} {\bibfnamefont {A.~S.}\ \bibnamefont
			{Holevo}},\ }\href@noop {} {\bibfield  {journal} {\bibinfo  {journal} {Probl.
				Peredachi Inf.}\ }\textbf {\bibinfo {volume} {9}},\ \bibinfo {pages} {3}
		(\bibinfo {year} {1973})},\ \bibinfo {note} {{E}nglish translation
		\emph{Problems Inform. Transmission}, vol. 9, no. 3, pp.177--183,
		1973}\BibitemShut {NoStop}%
	\bibitem [{\citenamefont {Cleve}\ \emph {et~al.}(1999)\citenamefont {Cleve},
		\citenamefont {van Dam}, \citenamefont {Nielsen},\ and\ \citenamefont
		{Tapp}}]{CDNT99}%
	\BibitemOpen
	\bibfield  {author} {\bibinfo {author} {\bibfnamefont {R.}~\bibnamefont
			{Cleve}}, \bibinfo {author} {\bibfnamefont {W.}~\bibnamefont {van Dam}},
		\bibinfo {author} {\bibfnamefont {M.}~\bibnamefont {Nielsen}}, \ and\
		\bibinfo {author} {\bibfnamefont {A.}~\bibnamefont {Tapp}},\ }in\ \href@noop
	{} {\emph {\bibinfo {booktitle} {Quantum Computing and Quantum
				Communications}}},\ \bibinfo {editor} {edited by\ \bibinfo {editor}
		{\bibfnamefont {C.~P.}\ \bibnamefont {Williams}}}\ (\bibinfo  {publisher}
	{Springer Berlin Heidelberg},\ \bibinfo {address} {Berlin, Heidelberg},\
	\bibinfo {year} {1999})\ pp.\ \bibinfo {pages} {61--74}\BibitemShut {NoStop}%
	\bibitem [{\citenamefont {Devetak}\ \emph {et~al.}(2004)\citenamefont
		{Devetak}, \citenamefont {Harrow},\ and\ \citenamefont {Winter}}]{DHW04}%
	\BibitemOpen
	\bibfield  {author} {\bibinfo {author} {\bibfnamefont {I.}~\bibnamefont
			{Devetak}}, \bibinfo {author} {\bibfnamefont {A.~W.}\ \bibnamefont {Harrow}},
		\ and\ \bibinfo {author} {\bibfnamefont {A.}~\bibnamefont {Winter}},\ }\href
	{\doibase 10.1103/PhysRevLett.93.230504} {\bibfield  {journal} {\bibinfo
			{journal} {Phys. Rev. Lett.}\ }\textbf {\bibinfo {volume} {93}},\ \bibinfo
		{pages} {230504} (\bibinfo {year} {2004})}\BibitemShut {NoStop}%
	\bibitem [{\citenamefont {Nayak}\ and\ \citenamefont {Salzman}(2006)}]{NS06}%
	\BibitemOpen
	\bibfield  {author} {\bibinfo {author} {\bibfnamefont {A.}~\bibnamefont
			{Nayak}}\ and\ \bibinfo {author} {\bibfnamefont {J.}~\bibnamefont
			{Salzman}},\ }\href@noop {} {\bibfield  {journal} {\bibinfo  {journal} {J.
				ACM}\ }\textbf {\bibinfo {volume} {53}},\ \bibinfo {pages} {184} (\bibinfo
		{year} {2006})}\BibitemShut {NoStop}%
	\bibitem [{\citenamefont {{Hsieh}}\ and\ \citenamefont {{Wilde}}(2010)}]{HW10}%
	\BibitemOpen
	\bibfield  {author} {\bibinfo {author} {\bibfnamefont {M.}~\bibnamefont
			{{Hsieh}}}\ and\ \bibinfo {author} {\bibfnamefont {M.~M.}\ \bibnamefont
			{{Wilde}}},\ }\href@noop {} {\bibfield  {journal} {\bibinfo  {journal} {IEEE
				Trans. Inf. Theory}\ }\textbf {\bibinfo {volume} {56}},\ \bibinfo {pages}
		{4705} (\bibinfo {year} {2010})}\BibitemShut {NoStop}%
	\bibitem [{\citenamefont {Datta}\ and\ \citenamefont {Hsieh}(2011)}]{DH11}%
	\BibitemOpen
	\bibfield  {author} {\bibinfo {author} {\bibfnamefont {N.}~\bibnamefont
			{Datta}}\ and\ \bibinfo {author} {\bibfnamefont {M.-H.}\ \bibnamefont
			{Hsieh}},\ }\href {\doibase 10.1088/1367-2630/13/9/093042} {\bibfield
		{journal} {\bibinfo  {journal} {New J. Phys.}\ }\textbf {\bibinfo {volume}
			{13}},\ \bibinfo {pages} {093042} (\bibinfo {year} {2011})}\BibitemShut
	{NoStop}%
	\bibitem [{\citenamefont {Lai}\ and\ \citenamefont {Chung}(2019)}]{LC18c}%
	\BibitemOpen
	\bibfield  {author} {\bibinfo {author} {\bibfnamefont {C.-Y.}\ \bibnamefont
			{Lai}}\ and\ \bibinfo {author} {\bibfnamefont {K.-M.}\ \bibnamefont
			{Chung}},\ }in\ \href@noop {} {\emph {\bibinfo {booktitle} {Proc. IEEE Intl.
				Symp. Inf. Theory}}}\ (\bibinfo {year} {2019})\ p.\ \bibinfo {pages} {2997},\
	\bibinfo {note} {arXiv:1809.10694}\BibitemShut {NoStop}%
	\bibitem [{\citenamefont {Paw\l{l}owski}\ \emph {et~al.}(2009)\citenamefont
		{Paw\l{l}owski}, \citenamefont {Paterek}, \citenamefont {Kaszlikowski},
		\citenamefont {Scarani}, \citenamefont {Winter},\ and\ \citenamefont
		{\.{Z}ukowski}}]{PPK+09}%
	\BibitemOpen
	\bibfield  {author} {\bibinfo {author} {\bibfnamefont {M.}~\bibnamefont
			{Paw\l{l}owski}}, \bibinfo {author} {\bibfnamefont {T.}~\bibnamefont
			{Paterek}}, \bibinfo {author} {\bibfnamefont {D.}~\bibnamefont
			{Kaszlikowski}}, \bibinfo {author} {\bibfnamefont {V.}~\bibnamefont
			{Scarani}}, \bibinfo {author} {\bibfnamefont {A.}~\bibnamefont {Winter}}, \
		and\ \bibinfo {author} {\bibfnamefont {M.}~\bibnamefont {\.{Z}ukowski}},\
	}\href@noop {} {\bibfield  {journal} {\bibinfo  {journal} {Nature}\ }\textbf
		{\bibinfo {volume} {461}},\ \bibinfo {pages} {1101–1104} (\bibinfo {year}
		{2009})}\BibitemShut {NoStop}%
	\bibitem [{\citenamefont {Almeida}\ \emph {et~al.}(2010)\citenamefont
		{Almeida}, \citenamefont {Bancal}, \citenamefont {Brunner}, \citenamefont
		{Ac\'{\i}n}, \citenamefont {Gisin},\ and\ \citenamefont {Pironio}}]{ABB+10}%
	\BibitemOpen
	\bibfield  {author} {\bibinfo {author} {\bibfnamefont {M.~L.}\ \bibnamefont
			{Almeida}}, \bibinfo {author} {\bibfnamefont {J.-D.}\ \bibnamefont {Bancal}},
		\bibinfo {author} {\bibfnamefont {N.}~\bibnamefont {Brunner}}, \bibinfo
		{author} {\bibfnamefont {A.}~\bibnamefont {Ac\'{\i}n}}, \bibinfo {author}
		{\bibfnamefont {N.}~\bibnamefont {Gisin}}, \ and\ \bibinfo {author}
		{\bibfnamefont {S.}~\bibnamefont {Pironio}},\ }\href@noop {} {\bibfield
		{journal} {\bibinfo  {journal} {Phys. Rev. Lett.}\ }\textbf {\bibinfo
			{volume} {104}},\ \bibinfo {pages} {230404} (\bibinfo {year}
		{2010})}\BibitemShut {NoStop}%
	\bibitem [{\citenamefont {Del~Santo}\ and\ \citenamefont
		{Daki\ifmmode~\acute{c}\else \'{c}\fi{}}(2018)}]{SD18}%
	\BibitemOpen
	\bibfield  {author} {\bibinfo {author} {\bibfnamefont {F.}~\bibnamefont
			{Del~Santo}}\ and\ \bibinfo {author} {\bibfnamefont {B.}~\bibnamefont
			{Daki\ifmmode~\acute{c}\else \'{c}\fi{}}},\ }\href@noop {} {\bibfield
		{journal} {\bibinfo  {journal} {Phys. Rev. Lett.}\ }\textbf {\bibinfo
			{volume} {120}},\ \bibinfo {pages} {060503} (\bibinfo {year}
		{2018})}\BibitemShut {NoStop}%
	\bibitem [{\citenamefont {{Massa}}\ \emph {et~al.}(2018)\citenamefont
		{{Massa}}, \citenamefont {{Moqanaki}}, \citenamefont {{Del Santo}},
		\citenamefont {{Dakic}},\ and\ \citenamefont {{Walther}}}]{MMD+18}%
	\BibitemOpen
	\bibfield  {author} {\bibinfo {author} {\bibfnamefont {F.}~\bibnamefont
			{{Massa}}}, \bibinfo {author} {\bibfnamefont {A.}~\bibnamefont {{Moqanaki}}},
		\bibinfo {author} {\bibfnamefont {F.}~\bibnamefont {{Del Santo}}}, \bibinfo
		{author} {\bibfnamefont {B.}~\bibnamefont {{Dakic}}}, \ and\ \bibinfo
		{author} {\bibfnamefont {P.}~\bibnamefont {{Walther}}},\ }in\ \href@noop {}
	{\emph {\bibinfo {booktitle} {2018 Conference on Lasers and Electro-Optics
				Pacific Rim (CLEO-PR)}}}\ (\bibinfo {year} {2018})\ pp.\ \bibinfo {pages}
	{1--2}\BibitemShut {NoStop}%
	\bibitem [{\citenamefont {Renninger}(1953)}]{Ren53}%
	\BibitemOpen
	\bibfield  {author} {\bibinfo {author} {\bibfnamefont {M.}~\bibnamefont
			{Renninger}},\ }\href@noop {} {\bibfield  {journal} {\bibinfo  {journal}
			{Zeitschrift für Physik}\ }\textbf {\bibinfo {volume} {136}},\ \bibinfo
		{pages} {251} (\bibinfo {year} {1953})}\BibitemShut {NoStop}%
	\bibitem [{\citenamefont {Baere}(2005)}]{Bae05}%
	\BibitemOpen
	\bibfield  {author} {\bibinfo {author} {\bibfnamefont {W.~D.}\ \bibnamefont
			{Baere}},\ }\href@noop {} {\  (\bibinfo {year} {2005})},\ \bibinfo {note}
	{arXiv:quant-ph/0504031}\BibitemShut {NoStop}%
	\bibitem [{\citenamefont {Elitzur}\ and\ \citenamefont
		{Vaidman}(1993)}]{Elitzur1993}%
	\BibitemOpen
	\bibfield  {author} {\bibinfo {author} {\bibfnamefont {A.~C.}\ \bibnamefont
			{Elitzur}}\ and\ \bibinfo {author} {\bibfnamefont {L.}~\bibnamefont
			{Vaidman}},\ }\href {\doibase 10.1007/BF00736012} {\bibfield  {journal}
		{\bibinfo  {journal} {Foundations of Physics}\ }\textbf {\bibinfo {volume}
			{23}},\ \bibinfo {pages} {987} (\bibinfo {year} {1993})}\BibitemShut
	{NoStop}%
	\bibitem [{\citenamefont {Horvat}\ and\ \citenamefont
		{Daki\'{c}}(2019)}]{HD19}%
	\BibitemOpen
	\bibfield  {author} {\bibinfo {author} {\bibfnamefont {S.}~\bibnamefont
			{Horvat}}\ and\ \bibinfo {author} {\bibfnamefont {B.}~\bibnamefont
			{Daki\'{c}}},\ }\href@noop {} {\  (\bibinfo {year} {2019})},\ \Eprint
	{http://arxiv.org/abs/arXiv:2003.12114} {arXiv:2003.12114} \BibitemShut
	{NoStop}%
	\bibitem [{\citenamefont {Senellart}\ \emph {et~al.}(2017)\citenamefont
		{Senellart}, \citenamefont {Solomon},\ and\ \citenamefont {White}}]{SSW17}%
	\BibitemOpen
	\bibfield  {author} {\bibinfo {author} {\bibfnamefont {P.}~\bibnamefont
			{Senellart}}, \bibinfo {author} {\bibfnamefont {G.}~\bibnamefont {Solomon}},
		\ and\ \bibinfo {author} {\bibfnamefont {A.}~\bibnamefont {White}},\
	}\href@noop {} {\bibfield  {journal} {\bibinfo  {journal} {Nature
				Nanotechnology}\ }\textbf {\bibinfo {volume} {12}},\ \bibinfo {pages} {1026}
		(\bibinfo {year} {2017})}\BibitemShut {NoStop}%
\end{thebibliography}
\end{document}